\documentclass[11pt,a4paper,oneside,english]{article}
\pdfoutput=1
\pdfoutput=1
\pdfoutput=1
\pdfoutput=1
\pdfoutput=1

\usepackage{graphicx}
\usepackage{amsmath,amssymb}
\usepackage{babel}

\usepackage[unicode,hidelinks,pdfauthor={T. Kohout},bookmarksopen=true,pdfpagemode=UseNone,pdfstartview={XYZ null null 1.0}]{hyperref}

\usepackage[numbers,sort&compress]{natbib}

\usepackage{lastpage}
\usepackage{fancyhdr}
\fancypagestyle{plain}{

\fancyhead[L]{}
\fancyhead[R]{}
\fancyhead[C]{\small T. Kohout et al. Insight into shock-induced changes in asteroid regoliths. Icarus.}
\fancyfoot{}
\fancyfoot[R]{\small page \thepage/\pageref{LastPage}}
}
\fancypagestyle{empty}{

\fancyhead{}
\fancyfoot{}
\fancyfoot[R]{\small page \thepage/\pageref{LastPage}}
}

\usepackage{textcomp}
\usepackage[utf8]{inputenc}
\usepackage[T1]{fontenc}
\usepackage{tgadventor} 
\usepackage{tgpagella} 
\usepackage[bf,sf]{titlesec}
\def \micron {{\textmu}m}

\newcommand{\degr}[0]{\ensuremath{{}^{\circ}}}
\newcommand{\mytimes}[0]{\ensuremath{\!\times\!}}

\usepackage{authblk}

\title{Mineralogy, reflectance spectra, and physical properties of the Chelyabinsk LL5 chondrite --- Insight into shock-induced changes in asteroid regoliths}

\author[1,2]{Tomas Kohout}
\author[3,4,5]{Maria Gritsevich}
\author[6]{Victor I. Grokhovsky}
\author[6]{Grigoriy A. Yakovlev}
\author[7,8]{Jakub Haloda}
\author[7]{Patricie Halodova}
\author[9]{Radoslaw M. Michallik}
\author[1]{Antti Penttil\"a}
\author[1,3]{Karri Muinonen}

\affil[1]{Department of Physics, University of Helsinki, P.O. Box 64, 00014 Helsinki University, Finland (tomas.kohout@helsinki.fi)}
\affil[2]{Institute of Geology, Academy of Sciences of the Czech Republic, Rozvojov\'a 269, 16500 Prague 6, Czech Republic}
\affil[3]{Finnish Geodetic Institute, Geodeetinrinne 2, P.O. Box 15, FI-02431 Masala, Finland}
\affil[4]{Institute of Mechanics, Lomonosov Moscow State University, Michurinsky prt., 1, 119192 Moscow, Russia}
\affil[5]{Russian Academy of Sciences, Dorodnicyn Computing Centre, Department of Computational Physics, Vavilova ul. 40, 119333 Moscow, Russia}
\affil[6]{Ural Federal University, Mira st., 620002 Ekaterinburg, Russia}
\affil[7]{Czech Geological Survey, Geologick\'a 6, 152 00 Praha 5, Czech Republic}
\affil[8]{Oxford Instruments NanoAnalysis, Halifax Road, High Wycombe, Bucks HP12 3SE, United Kingdom}
\affil[9]{Department of Geosciences and Geography, University of Helsinki, P.O. Box 64, 00014 Helsinki University, Finland}

\begin{document}

\maketitle
\thispagestyle{empty}

\hspace{10mm}\\
Cite as: T. Kohout et al. Mineralogy, reflectance spectra, and physical properties of the Chelyabinsk LL5 chondrite --- Insight into shock induced changes in asteroid regoliths. Icarus {\bf 228}, 78--85, 2014, DOI:10.1016/j.icarus.2013.09.027.

\newpage
\pagestyle{plain}

\begin{abstract}
\noindent The mineralogy and physical properties of Chelyabinsk meteorites (fall, February 15, 2013) are presented. Three types of meteorite material are present, described as the light-colored, dark-colored, and impact-melt lithologies. All are of LL5 composition with the impact-melt lithology being close to whole-rock melt and the dark-colored lithology being shock-darkened due to partial melting of iron metal and sulfides. This enables us to study the effect of increasing shock on material with identical composition and origin. Based on the magnetic susceptibility, the Chelyabinsk meteorites are richer in metallic iron as compared to other LL chondrites. The measured bulk and grain densities and the porosity closely resemble other LL chondrites. Shock darkening does not have a significant effect on the material physical properties, but causes a decrease of reflectance and decrease in silicate absorption bands in the reflectance spectra. This is similar to the space weathering effects observed on asteroids. However, compared to space weathered materials, there is a negligible to minor slope change observed in impact-melt and shock-darkened meteorite spectra. Thus, it is possible that some dark asteroids with invisible silicate absorption bands may be composed of relatively fresh shock-darkened chondritic material.
\end{abstract}

\section*{Introduction}

On February 15, 2013, at 9:22 am, an exceptionally bright and long-duration fireball was observed by many eyewitnesses in the Chelyabinsk region, Russia \cite{metbul,galimov-2013}. A large-sized object with a relatively low mass-loss rate and shallow atmospheric entry angle led to a very long trail. The fireball was observed in the Chelyabinsk, Kurgan, Orenburg, Tyumen, Ekaterinburg, Kostanay, and Aktobe regions, and in the Republic of Bashkortostan, Russia, and in Kazakhstan. The event was recorded by numerous video cameras from the ground and was also imaged from space by the Meteosat and Fengyun satellites \cite{proud-2013}. A strong shock wave associated with the fireball caused significant damage including broken windows and partial building collapses in Chelyabinsk and the surrounding territories.

\begin{figure}[hbt]
\centering
\includegraphics{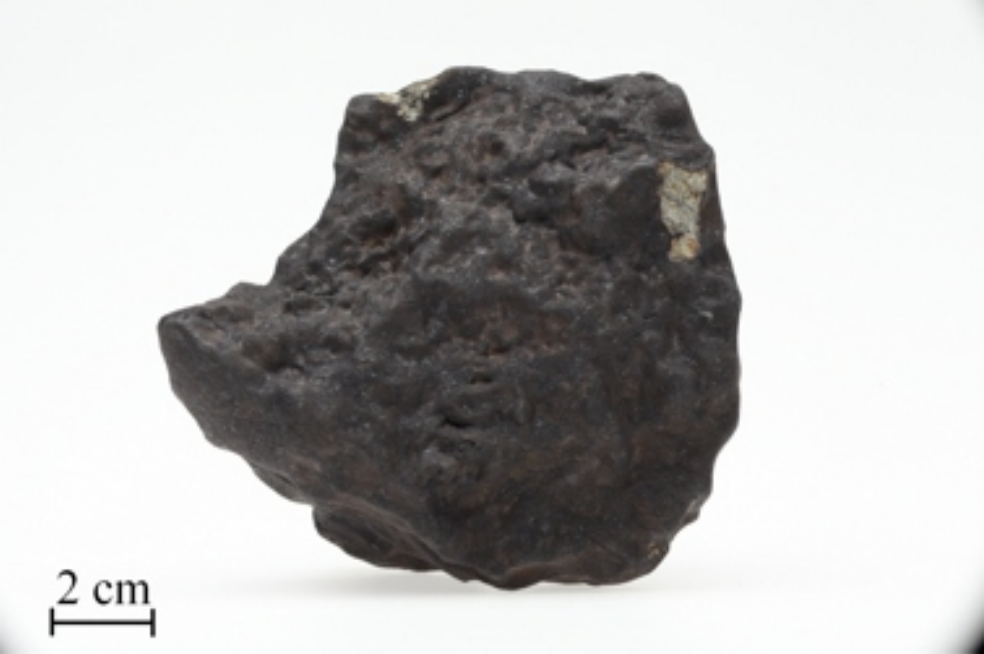}
\caption{1.8 kg A50 meteorite is the largest piece in the Ural Federal University collections.}
\label{fig:one}
\end{figure}

Two days later the first fragments of the Chelyabinsk meteorite were reported to be found around Pervomaiskoe, Deputatsky, and Yemanzhelinka, located approximately 40 km south of Chelyabinsk. Successful search campaigns in this area were organized by the scientists from the Ural Federal University as well as by the Laboratory of Meteoritics of the Vernadsky Institute, Russian Academy of Science. Numerous meteorite samples were also recovered and collected out of snow by local residents before a subsequent extensive snowfall. The majority of fragments are composed of relatively small pieces, with the largest officially reported being 3.4 kg as of this publication. The largest piece in the Ural Federal University collections is a 1.8 kg sample no. A50 (Fig.~\ref{fig:one}). A large amount of additional material was collected on the ground along the fireball track after the snow melted at the end of April and in May 2013. The total collected mass of the Chelyabinsk meteorites exceeds 100 kg.

\begin{figure}[hbt]
\centering
\includegraphics{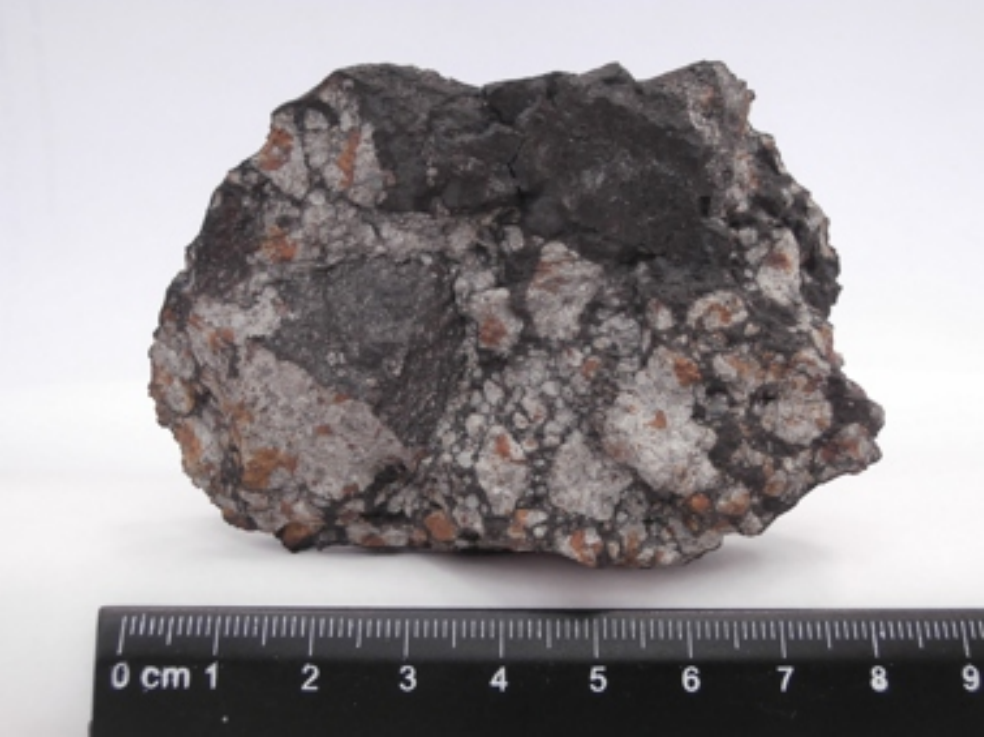}
\caption{310.6 g A2 meteorite is a breccia of light-colored lithology clasts within impact-melt lithology.}
\label{fig:two}
\end{figure}

As more meteorite samples were collected, it became apparent that multiple types of material are present. The recovered meteorites are composed of either light-colored or dark-colored (shock-darkened) lithology \cite{metbul}. Additionally, dark impact-melt is being present in varying amounts within light- and dark-colored stones. Impact-melt is distinct in appearance from the dark-colored (shock-darkened) lithology and is distinguished as a third lithology type. Some meteorites contain significant portion of this impact-melt lithology (Fig.~\ref{fig:two}). However, the light- and dark-colored (shock-darkened) lithologies were (up to our knowledge) not found together within one meteorite.

According to the computations by Yeomans and Chodas \cite{yeomans-2013}, the initial diameter of the meteoroid (or a small asteroid) was approximately 17--20 m with a mass of approximately $11 \mytimes 10^6$ kg. This object is the largest observed celestial body colliding with the Earth since the Tunguska event in 1908.

In this study, we report physical properties (bulk and grain density, porosity, and magnetic susceptibility) of 49 individual Chelyabinsk meteorites from the Ural Federal University collections, Ekaterinburg, Russian Federation. Detailed mineralogical and spectral characterization of two samples (representing the light-colored and dark-colored lithologies, both with impact-melt veins) is presented to determine and explain the differences between the lithologies. Finally, the origin of the Chelyabinsk meteorites and their parent body history are briefly discussed.

\section*{Materials and methods}

\subsection*{Mineralogy and chemistry}

The mineral composition was determined at the Department of Geosciences and Geography, University of Helsinki, with a JEOL JXA-8600 Superprobe using standard-calibrated energy-dispersive spectrometry (EDS) on a nitrogen-cooled Si(Li) detector. Two samples (the light-colored and dark-colored lithologies) were polished and carbon-coated prior to the measurements. Matrix correction was performed with the XMas software using the PAP correction method. A combination of natural and synthetic silicate and oxide standards was used for calibration. The accuracy was determined by cross-analyses of standard materials and was better than 2\% for the major elements (>10 wt\%) and better than 5\% for the minor elements (1--10 wt\%). Oxygen was calculated by stoichiometry. The analysis conditions were 15 kV accelerating voltage, 1 nA beam current and 100 s acquisition time per spot with a focused beam. Altogether, 116 spots were analyzed, 58 in the light-colored lithology sample (no. VG1) and 58 in the dark-colored lithology sample (no. VG4b).

Because only major and minor elements were analyzed, the EDS method was considered accurate enough and faster than the commonly used more accurate wavelength-dispersive spectrometry (WDS) method in quantitative microanalysis.

Subsequently, two polished thin sections from these two samples (27 mm $\times$ 15 mm from the light-colored lithology sample no. VG1 and 12 mm $\times$ 14 mm from the dark-colored lithology sample no. VG4b) were prepared. Texture, mineralogical characteristics, and shock features were studied using a LEICA DMLP petrographic microscope at the Czech Geological Survey.

\subsection*{Spectral measurements}

The spectral characterization was carried out using an OL 750 automated spectroradiometric measurement system by Gooch \& Housego at the Department of Physics, University of Helsinki. The spectroradiometer is equipped with a quartz tungsten-halogen light source, a monochromator, an integrating sphere with a polytetrafluoroethylene (PTFE) coating, and two high sensitivity detectors.

The incident beam has an angle of 10\degr{} with the sample holder normal vector. There is a light trap for specular reflection at the mirror angle. The detector is positioned so that there is no direct reflection from the sample to the detector and all the detected signal is multiply scattered in the PTFE coated integrating sphere. The system is closed during the measurement and there is no stray light.

The detectors use either a silicon detector head (for wavelengths from 0.2--1.1 \micron) or a cooled PbS detector head (for wavelengths up to 3.2 \micron).

In the diffuse spectral measurement procedure with the spectroradiometer, the spectra of the PTFE coating in the integrating sphere is always measured before the actual target while the target is already placed in the sample holder. By comparing these reflectances and correcting with the known calibration values of the PTFE material, the reflectance spectra, normalized against an ideal diffuser, are obtained.

The reflectance of the light-colored (sample no. VG1), dark-colored (sample no. VG4b), and impact-melt (vein in the sample no. M18) lithologies was acquired in the wavelength range of 500--2500 nm from saw-cut (VG1 and VG4b) or broken (M18) surfaces. The surfaces were rough enough preventing a significant direct specular reflection and were identical to those used for mineralogical characterization. The size of the incident light beam footprint on the target can be varied using slits and apertures in the monochromator. The light-colored lithology sample (no. VG1) had a slightly larger surface area available for the spectral measurement, and therefore a 5 mm slit and a 5 mm aperture were used for this target. Thus, the resulting collimated beam footprint on the target had a diameter of 5 mm. The dark-colored lithology sample (no. VG4b) and impact-melt lithology (vein in sample no. M18) were measured with a 2.5 mm slit and a 3 mm aperture, and the beam footprint on the target was approximately 2.5 mm.

\subsection*{Density and porosity}

The density and porosity values were obtained using the mobile laboratory described in \cite{kohout-2008} deployed at the Ural Federal University, Ekaterinburg. The bulk volume was determined using a modified Archimedean method \cite{consolmagno-1998,macke-2010} incorporating glass beads $\sim\!\!0.3$ mm in diameter. Ten sets of measurements per sample were carried out. The method was thoroughly tested and calibrated prior to the measurements using volume standards and the absolute resolution and precision were determined to be $\pm 0.1$ cm$^3$. The grain volume of the meteorites was measured using a Quantachrome Ultrapyc 1300e He pycnometer. The absolute resolution and the precision of this device are better than $\pm 0.01$ cm$^3$. Thus, the relative errors of both volumetric methods increase with decreasing sample size. However, compared to the bulk volume, the grain volume is determined with higher resolution and precision. Meteorite masses were determined using a digital OHAUS Navigator scale with 0.1 g resolution and precision. The scale was calibrated prior to the measurements using internal calibration.

\subsection*{Magnetic susceptibility}

The magnetic susceptibility of the samples smaller than 2.5 cm was measured using a ZH instruments SM-100 susceptibility meter operating at 8 kHz frequency and 320 A/m RMS field amplitude. For larger samples, a ZH instruments SM-30 portable surface susceptibility meter with a large 5 cm coil operating at the same 8 kHz frequency was used. The values were corrected for the actual sample size following the procedure described in Gattacceca et al. \cite{gattacceca-2004}.

The susceptibility of the samples was measured three times along three perpendicular directions. Subsequently, a logarithm of the average apparent magnetic susceptibility (in $10^{-9}$ m$^3$/kg) was calculated as described in Rochette et al. \cite{rochette-2003}. The relative error in the determined value of the magnetic susceptibility logarithm is below 3\%.

For the statistical calculations of the density, porosity, and susceptibility, only samples larger than 5 g were considered to be homogeneous enough to be representative of the whole-rock material.

\section*{Results}

\subsection*{Mineralogy and shock characterization}

\begin{figure}[!hbt]
\centering
\includegraphics{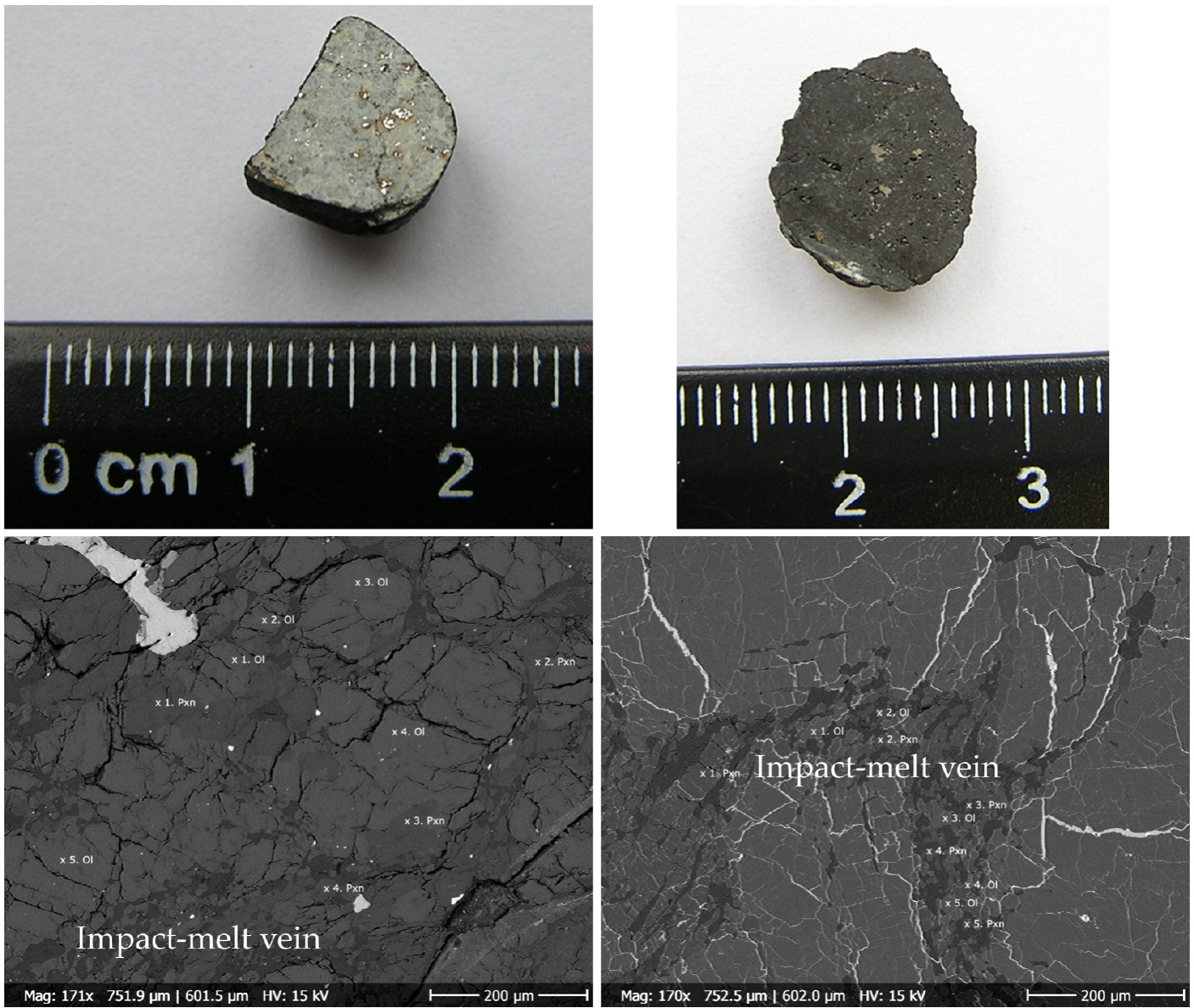}
\caption{Overall optical (up) and high-resolution scanning electron microscope images in back-scattered electrons (SEM-BSE, down) of the light-colored (no. VG1, left) and dark-colored (no. VG4b, right) lithologies. In the SEM-BSE images, coarse-grained clasts made of olivine and pyroxene surrounded by fine-grained impact-melt lithology veins can be identified in both lithologies. The abundant internal grain fracturing can be seen as darker voids. In the dark-colored lithology, the intra- and inter-granular fractures are filled with sulphide and metal-rich impact-melt (bright material in the SEM-BSE images). Points marked as Ol and Pxn indicate grains with determined composition by energy-dispersive spectrometry (EDS).}
\label{fig:three}
\end{figure}

\begin{figure}[!hbt]
\centering
\includegraphics{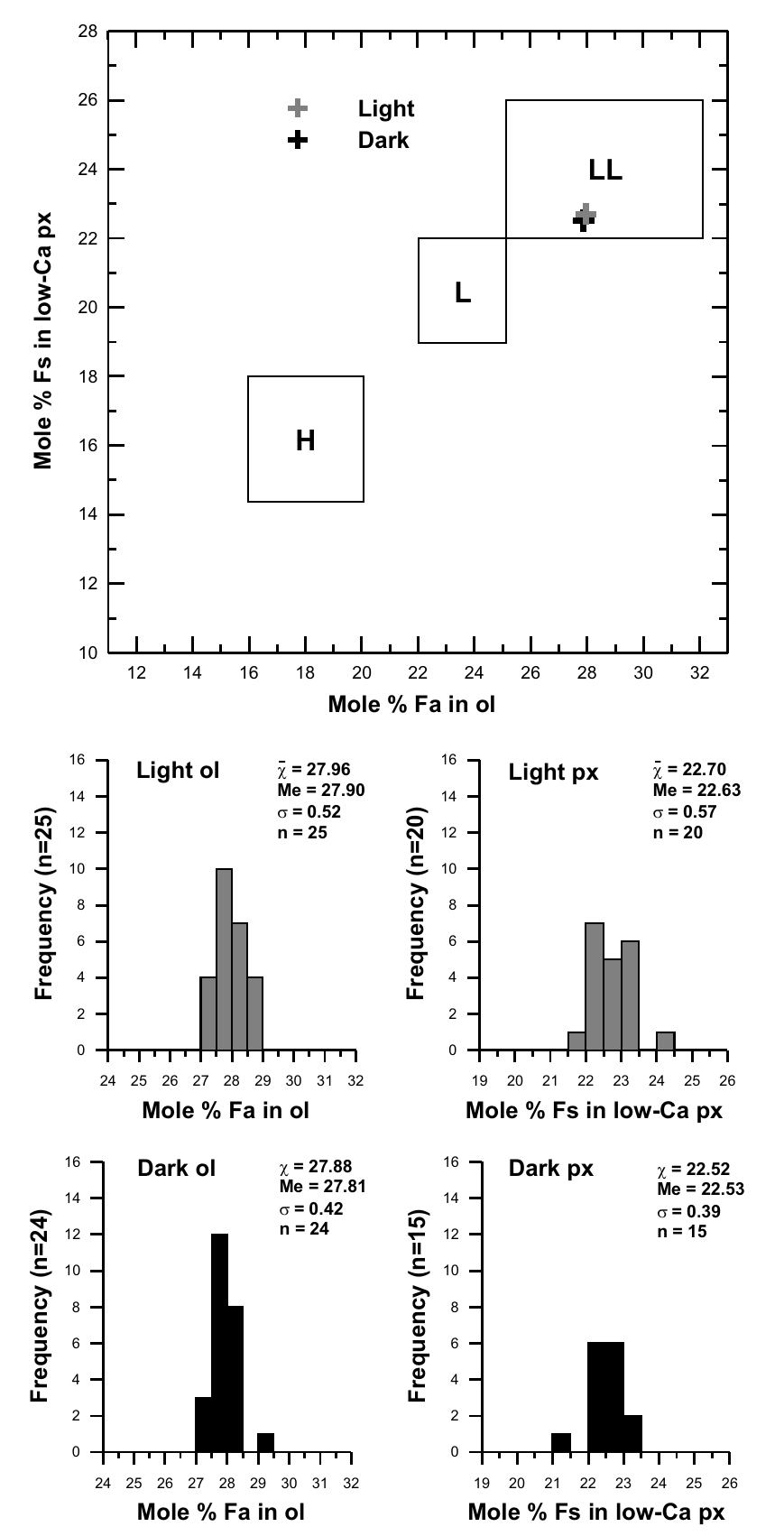}
\caption{Mineralogical classification of the olivine (ol) and pyroxene (px) clasts based on the molar orthopyroxene ferrosilite (Fs) and olivine fayalite (Fa) amount. H, L, and LL indicate typical ranges for H, L, and LL ordinary chondrites compiled from Brearley and Jones \cite{brearley-1998}. $x$ -- average, $Me$ -- median, $\sigma$ -- standard deviation, $n$ -- number of analyzed spots. }
\label{fig:four}
\end{figure}

Figure~\ref{fig:three} shows the overall optical and detailed SEM images of the light-colored (no. VG1) and dark-colored (no. VG4b) lithology meteorites. The computed olivine fayalite (Fa) content and orthopyroxene ferrosilite (Fs) content determined by means of EDS (Fig.~\ref{fig:four}) from the main meteorite compounds (matrix and chondrules, excluding melt veins) reveal that the light-colored and dark-colored lithologies are almost identical in composition and, based on the classification scheme in Brearley and Jones (\cite{brearley-1998}, p. 3--8), can be classified as LL ordinary chondrites. The homogeneity of olivine and pyroxene as indicated by a standard deviation of the Fa and Fs compositions (standard deviation $\leq$ 5\%), texture showing recrystallized matrix with readily distinguished chondrules, and an absence of igneous glass in the chondrules are typical for petrographic type 5. Abundant impact-melt veins are found in both light-colored and dark-colored lithologies. Their silicate composition is almost identical to the surrounding meteorite material pointing again on their common source.

As revealed by optical microscopy and SEM imaging, the difference between the light-colored and dark-colored lithologies is in the level of shock. The light-colored lithology is slightly fractured and represents monomict LL5 breccia (Fig.~\ref{fig:three}). Individual clasts show slightly different shock features indicating a different intensity of shock metamorphism. The shock features observed in olivine for individual clasts vary from undulose extinction to strong mosaicism which is indicative of shock levels S3 and S4 \cite{stoffler-1991}. The silicate grains show microscopic fracturing. The large inter-granular cracks are filled with dark melt veins containing also fine-grained mineral fragments. The silicate composition of this melt is close to that of the surrounding material. Thus, we interpret this melt as an impact-melt originating from the light-colored lithology.

The dark-colored lithology is shocked to a larger extent. The material is extensively fractured and abundant impact-melt is present (Fig.~\ref{fig:three}). Two populations of melt exist within the dark-colored lithology. The inter-granular space is filled with impact-melt veins containing fine-grained mineral fragments (first population) similar to impact-melt observed in the light-colored lithology. The silicate composition of these melt veins also closely resembles the main meteorite material with the additional presence of clinopyroxene. Additionally, the inter-granular as well as intra-granular fractures are impregnated with a dense network of thin, fine-grained melt veins composed predominantly of troilite and metal (the second melt population). This metallic melt is abundant through the dark-colored lithology and is the main darkening agent responsible for its optical appearance. The abovementioned findings are similar to these in independent study by Galimov et al. \cite{galimov-2013}.

\subsection*{Spectral measurements}

\begin{figure}[hbt]
\centering
\includegraphics{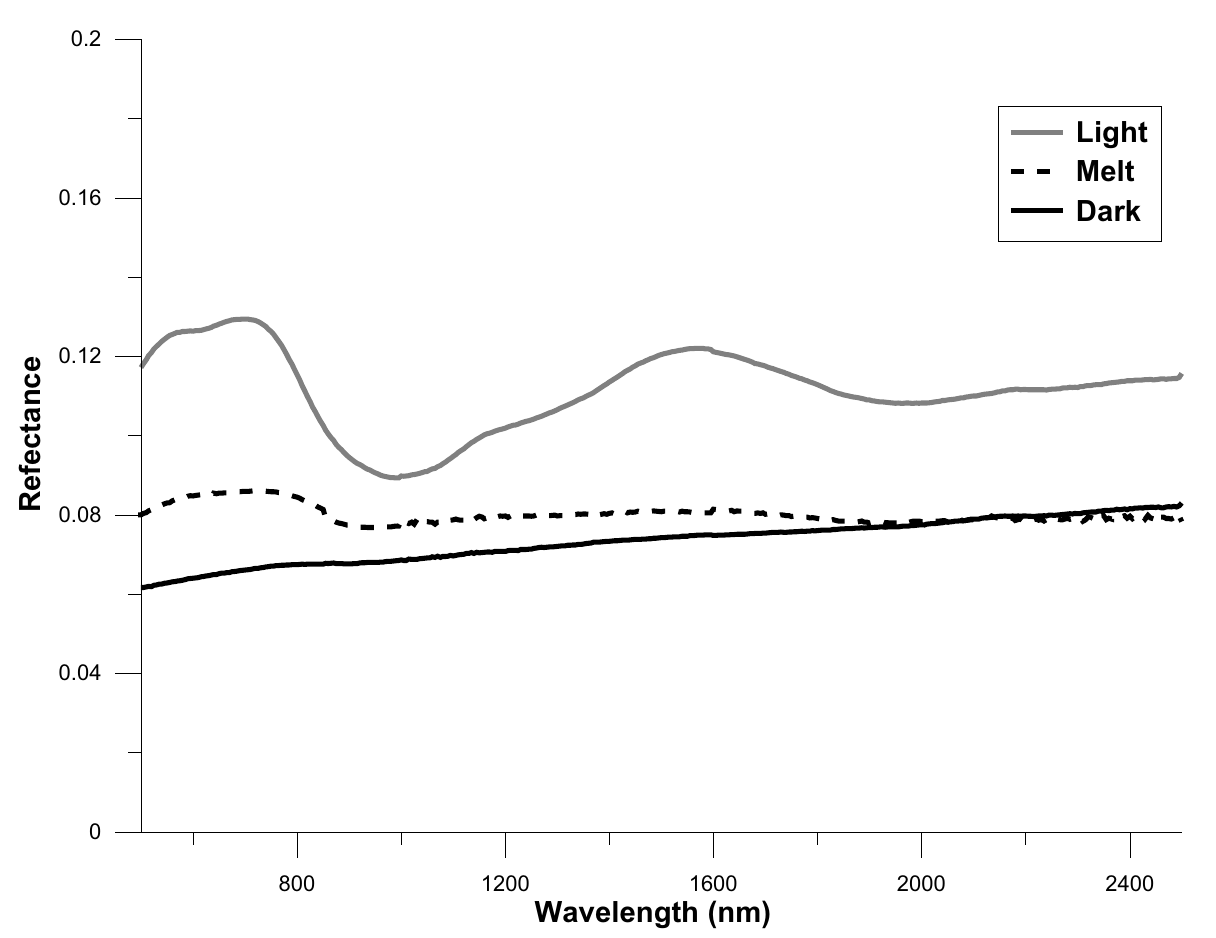}
\caption{Reflectance spectra of the light-colored, dark-colored, and impact-melt lithologies.}
\label{fig:five}
\end{figure}

The spectrum of the light-colored lithology shows broad 1 and 2 \micron{} absorption bands of olivine and pyroxene (Fig.~\ref{fig:five}). The overall normalized reflectance is between 0.10 and 0.14. The reflectance of the dark-colored lithology is lower (reflectance in the range of 0.06--0.08) and the 1 and 2 \micron{} absorption bands are almost indistinguishable in the spectrum. The reflectance of the impact-melt is being intermediate between the light-colored and dark-colored lithologies.

\subsection*{Density and porosity}

The results of the density and porosity measurements of the Chelyabinsk meteorites are summarized in Tables~\ref{tab:one} and \ref{tab:two}. The bulk density of the individual meteorites ranges from 3.14 to 3.48 g/cm$^3$ with the mean value of 3.32 g/cm$^3$ and with s.d. (standard deviation) of 0.09 g/cm$^3$ (light-colored lithology, 18 samples included), and 3.27 g/cm$^3$ s.d. 0.08 g/cm$^3$ (dark-colored, 7 samples included). The grain density ranges from 3.22 to 3.60 g/cm$^3$ with the mean value of 3.51 g/cm$^3$ s.d. 0.07 g/cm$^3$ (light-colored, 14 samples included) and 3.42 g/cm$^3$ s.d. 0.10 g/cm$^3$ (dark-colored, 7 samples included). The dark-colored lithology appears slightly less dense. However, the light-colored and dark-colored meteorite density ranges significantly overlap and neither bulk nor grain density can be used as a reliable criterion to distinguish between the two lithologies. Porosity is almost identical for both lithologies ranging from 1.5 to 11.4\% with the mean value of 6.0\% s.d. 3.2\% (light-colored, 13 samples included) and 5.7\% s.d. 1.7\% (dark-colored, 6 samples included).

\begin{table}[hbt]
\centering
\caption{List of Chelyabinsk meteorites included in the study with their measured physical properties. Meteorite no. corresponds to the catalogue of the Ural Federal University collections. For the statistical calculations of the density, porosity, and susceptibility, only samples larger than 5 g were considered homogeneous enough to be representative of the whole-rock material.}
\label{tab:one}
\vspace{1ex}
{\tiny
\begin{tabular}{l l p{5.5em} l l l l l l l l l l}
Meteorite & Date &  & Mass & \multicolumn{3}{c}{Dimensions (cm)} & $\rho_B$ & $\rho_G$ & $p$ & $\kappa_{mA}$ & $\kappa_V$ & log $\kappa_{mA}$ \\
no. & found & Litology & (g) & A & B & C & $(\frac{\mathrm{g}}{\mathrm{cm}^3})$ & $\frac{\mathrm{g}}{\mathrm{cm}^3})$ & (\%) & $(10^{-5} \frac{\mathrm{m}^3}{\mathrm{kg}})$ & $(10^{-6} \mathrm{SI})$ & $(\mathrm{in} 10^{-9} \frac{\mathrm{m}^3}{\mathrm{kg}})$ \\
\hline
VG3a & 22-23.2.2013 & light & 0.82 & 1.0 &  &  &  &  &  & 41.15 &  & 4.61 \\
VG3b & 22-23.2.2013 & light & 0.99 & 1.0 &  &  &  &  &  & 37.03 &  & 4.57 \\
VG2b & 22-23.2.2013 & light & 1.26 &  &  &  &  &  &  & 27.39 &  & 4.44 \\
VG2a & 22-23.2.2013 & light & 5.15 & 2.0 & 1.3 & 1.0 &  & 3.31 &  & 26.45 &  & 4.42 \\
A11a & 22-23.2.2013 & light & 8.23 & 3.2 & 1.5 & 1.4 & 3.38 & 3.43 & 2 & 30.76 & 104013 & 4.49 \\
U16 & 19.2.2013 & light & 12.84 & 2.0 & 2.0 & 1.5 & 3.29 & 3.46 & 5 & 27.31 & 89913 & 4.44 \\
A26 & 22-23.2.2013 & light & 15.38 & 2.8 & 2.8 & 2.0 & 3.35 & 3.51 & 5 & 35.42 & 118760 & 4.55 \\
M22 & 22-23.2.2013 & light & 15.66 & 3.0 & 2.4 & 1.8 & 3.40 & 3.60 & 5 & 30.74 & 104635 & 4.49 \\
Cz5 & 19.2.2013 & light & 18.97 & 2.8 & 2.5 & 1.8 & 3.39 & 3.54 & 4 & 38.29 & 129687 & 4.58 \\
U14 & 19.2.2013 & light & 19.50 & 3.2 & 2.1 & 1.9 & 3.15 & 3.53 & 11 & 33.16 & 104294 & 4.52 \\
U19 & 19.2.2013 & light & 20.51 & 2.5 & 2.3 & 2.0 & 3.26 & 3.58 & 9 & 23.37 & 76082 & 4.37 \\
A5 & 22-23.2.2013 & light & 29.93 & 3.5 & 2.4 & 1.8 & 3.48 & 3.54 & 2 & 33.00 & 114855 & 4.52 \\
KP1 & 21.4.2013 & light & 31.93 & 3.5 & 2.7 & 2.0 & 3.19 & 3.55 & 10 & 29.89 & 95440 & 4.48 \\
U13 & 19.2.2013 & light & 66.76 & 4.6 & 3.8 & 2.5 & 3.35 & 3.52 & 5 & 43.46 & 145798 & 4.64 \\
KP6 & 21.4.2013 & light & 142.70 & 5.0 & 4.3 & 3.8 & 3.39 & 3.51 & 3 & 27.54 & 93391 & 4.44 \\
A48 & 22-23.2.2013 & light & 147.10 & 9.0 & 7.0 & 2.0 &  &  &  &  & 77000 &  \\
TK1 & 27.4.2013 & light & 219.50 & 6.7 & 5.4 & 4.0 & 3.37 &  &  &  &  & 4.51 \\
U24 & 19.2.2013 & light & 273.30 & 8.0 & 5.5 & 5.0 & 3.30 &  &  & 31.62 & 104504 & 4.50 \\
MG1 & 27.4.2013 & light & 281.80 & 9.3 & 5.4 & 5.0 & 3.32 &  &  &  &  & 4.48 \\
KP7 & 21.4.2013 & light & 299.50 & 9.0 & 5.5 & 5.0 & 3.35 &  &  & 32.36 & 108312 & 4.51 \\
KP4 & 21.4.2013 & light breccia & 22.68 & 3.7 & 2.2 & 1.9 & 3.15 & 3.56 & 11 & 38.53 & 121387 & 4.59 \\
U9 & 19.2.2013 & light breccia & 77.92 & 5.5 & 4.0 & 3.5 & 3.27 & 3.47 & 6 & 27.54 & 90177 & 4.44 \\
A2 & 22-23.2.2013 & light breccia with melt & 310.60 & 8.0 & 6.0 & 4.5 & 3.28 &  &  & 29.51 & 96795 & 4.47 \\
VG1 & 22-23.2.2013 & light slice & 4.35 & 2.5 & 1.5 & 0.7 &  & 3.31 &  & 30.58 &  & 4.49 \\
VG4a & 22-23.2.2013 & dark & 0.90 & 1.3 &  &  &  &  &  & 41.37 &  & 4.62 \\
VG5a & 22-23.2.2013 & dark & 1.16 & 1.4 &  &  &  &  &  & 45.70 &  & 4.66 \\
VG5b & 22-23.2.2013 & dark & 1.17 & 1.2 &  &  &  &  &  & 44.10 &  & 4.64 \\
VG6b & 22-23.2.2013 & dark & 1.30 & 1.3 &  &  &  &  &  & 33.16 &  & 4.52 \\
VG6a & 22-23.2.2013 & dark & 1.76 & 1.3 &  &  &  &  &  & 32.34 &  & 4.51 \\
VG4b & 22-23.2.2013 & dark & 2.28 & 1.5 &  &  &  &  &  & 43.68 &  & 4.64 \\
KP2 & 21.4.2013 & dark & 7.52 & 2.2 & 1.7 & 1.4 &  & 3.22 &  & 21.17 &  & 4.33 \\
U2 & 19.2.2013 & dark & 11.61 & 2.5 & 2.0 & 1.0 & 3.14 & 3.39 & 8 & 44.10 & 138371 & 4.64 \\
U11 & 19.2.2013 & dark & 15.26 & 3.1 & 2.0 & 1.7 & 3.18 & 3.46 & 8 & 30.58 & 97249 & 4.49 \\
A6 & 22-23.2.2013 & dark & 18.89 & 2.8 & 2.5 & 2.0 & 3.31 & 3.50 & 5 & 41.32 & 136913 & 4.62 \\
A21 & 22-23.2.2013 & dark & 32.84 & 3.6 & 3.0 & 2.5 & 3.39 & 3.55 & 5 & 62.34 & 211082 & 4.79 \\
A20 & 22-23.2.2013 & dark & 39.15 & 3.5 & 3.3 & 3.0 & 3.26 & 3.44 & 5 & 36.95 & 120560 & 4.57 \\
MG2 & 27.4.2013 & dark & 60.20 & 4.6 & 4.2 & 2.7 & 3.32 &  &  &  &  & 4.40 \\
A3 & 22-23.2.2013 & dark & 75.28 & 5.5 & 4.0 & 3.5 & 3.26 & 3.37 & 3 & 21.38 & 69675 & 4.33 \\
M16 & 22-23.2.2013 & unclear & 6.54 & 2.0 & 1.5 & 1.2 &  & 3.37 &  & 32.88 &  & 4.52 \\
M26 & 22-23.2.2013 & unclear & 9.32 & 2.5 & 1.8 & 1.5 & 3.22 & 3.45 & 7 & 29.88 & 96065 & 4.48 \\
A9 & 22-23.2.2013 & unclear & 11.33 & 3.0 & 2.0 & 1.0 & 3.30 & 3.40 & 3 & 31.22 & 103016 & 4.49 \\
Cz1 & 19.2.2013 & unclear & 11.52 & 2.5 & 2.0 & 1.0 & 3.36 & 3.45 & 3 & 30.65 & 102993 & 4.49 \\
M1 & 22-23.2.2013 & unclear & 13.92 & 2.5 & 2.0 & 1.5 & 3.16 & 3.43 & 8 & 29.32 & 92733 & 4.47 \\
KP3 & 21.4.2013 & unclear & 16.77 & 3.1 & 2.1 & 1.4 & 3.24 & 3.58 & 10 & 22.46 & 72698 & 4.35 \\
M14 & 22-23.2.2013 & unclear & 20.25 & 3.8 & 2.4 & 1.6 & 3.29 & 3.52 & 7 & 32.80 & 107780 & 4.52 \\
Cz3 & 19.2.2013 & unclear & 30.20 & 3.0 & 2.6 & 2.5 & 3.18 & 3.56 & 11 & 23.49 & 74674 & 4.37 \\
KP5 & 21.4.2013 & unclear & 35.44 & 4.4 & 3.4 & 1.3 & 3.41 & 3.46 & 2 & 49.28 & 167936 & 4.69 \\
A49 & 22-23.2.2013 & unclear & 62.84 & 5.0 & 3.5 & 3.5 & 3.38 & 3.46 & 2 & 37.78 & 127638 & 4.58 \\
A47 & 22-23.2.2013 & unclear & 150.70 & 5.0 & 4.0 & 3.5 & 3.23 & 3.53 & 8 & 18.62 & 60089 & 4.27 \\
\end{tabular}
}
\end{table}

\begin{table}[hbt]
\centering
\caption{Comparison of the mean bulk and grain density, porosity, and magnetic susceptibility together with their standard deviations (s.d.) of the Chelyabinsk light-colored and dark-colored lithologies.}
\label{tab:two}
\vspace{1ex}
\begin{tabular}{p{4.2cm} l l}
 & Light-colored lithology & Dark-colored lithology \\
\hline
Bulk density (g/cm$^3$) & 3.32 (s.d. 0.09) & 3.27 (s.d.0.08) \\
Grain density (g/cm$^3$) & 3.51 (s.d. 0.07) & 3.42 (s.d. 0.10) \\
Porosity (\%) & 6.0 (s.d. 3.2) & 5.7 (s.d. 1.7) \\
Magnetic susceptibility (log in 10$^{-9}$ m$^3$/kg) & 4.49 (s.d. 0.07) & 4.52 (s.d. 0.15) \\
\end{tabular}
\end{table}

\begin{figure}[hbt]
\centering
\includegraphics{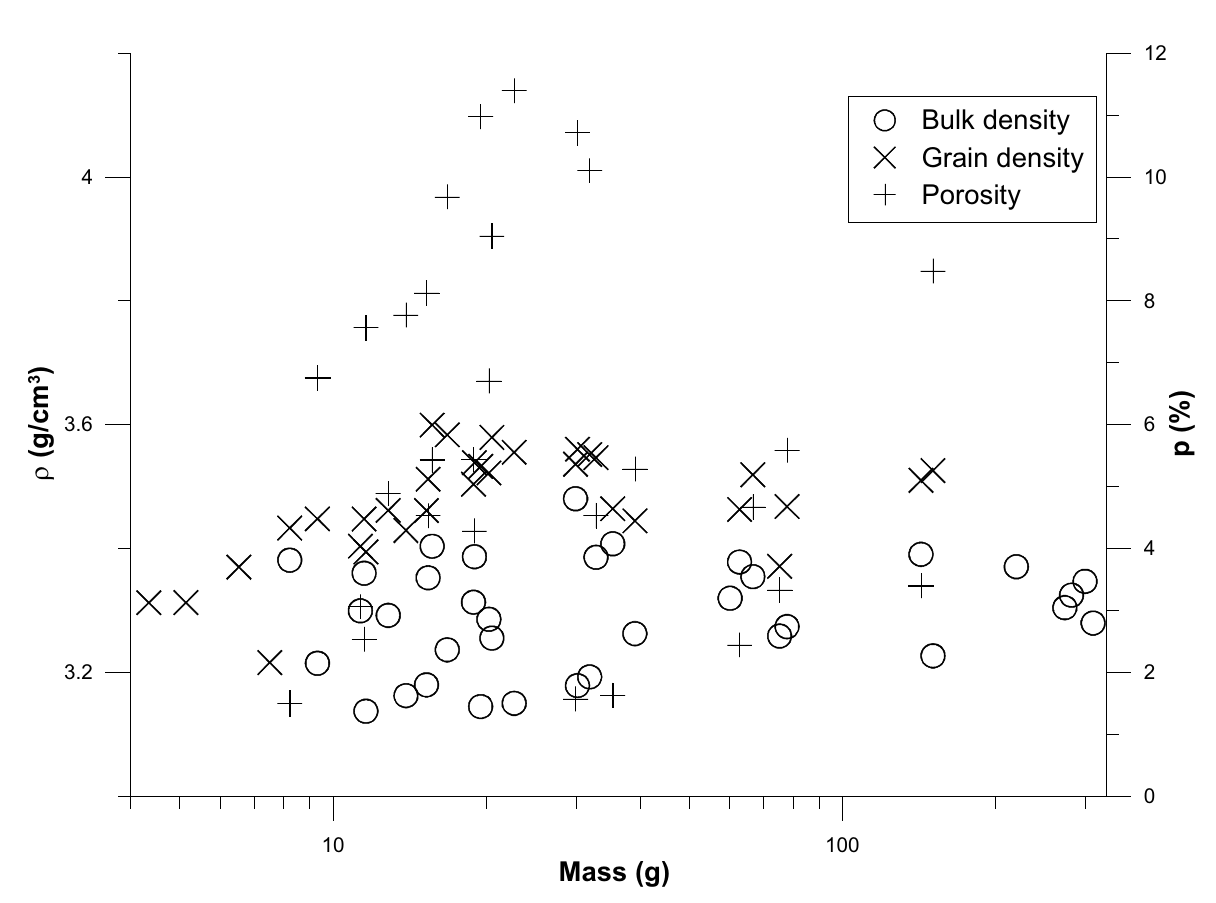}
\caption{Bulk and grain density ($\rho$) and porosity ($p$) of the Chelyabinsk meteorites as a function of their mass. No correlation is observed.}
\label{fig:six}
\end{figure}

The measured physical properties are in good agreement with the bulk (average 3.22 g/cm$^3$ s.d. 0.22 g/cm$^3$) and grain (3.54 g/cm$^3$ s.d. 0.13 g/cm$^3$) density and porosity (8.2\% s.d. 5.5\%) of other LL chondrite falls reported in Consolmagno et al. \cite{consolmagno-2008} and no correlation with the meteorite masses is observed (Fig.~\ref{fig:six}). Smaller samples show higher density and porosity scatter. This is most likely due to a higher relative uncertainty in the measured volume values of the smaller samples (as described in the Materials and methods section) and increasing inhomogeneity of the material at smaller scales.

\subsection*{Magnetic susceptibility}

\begin{figure}[hbt]
\centering
\includegraphics{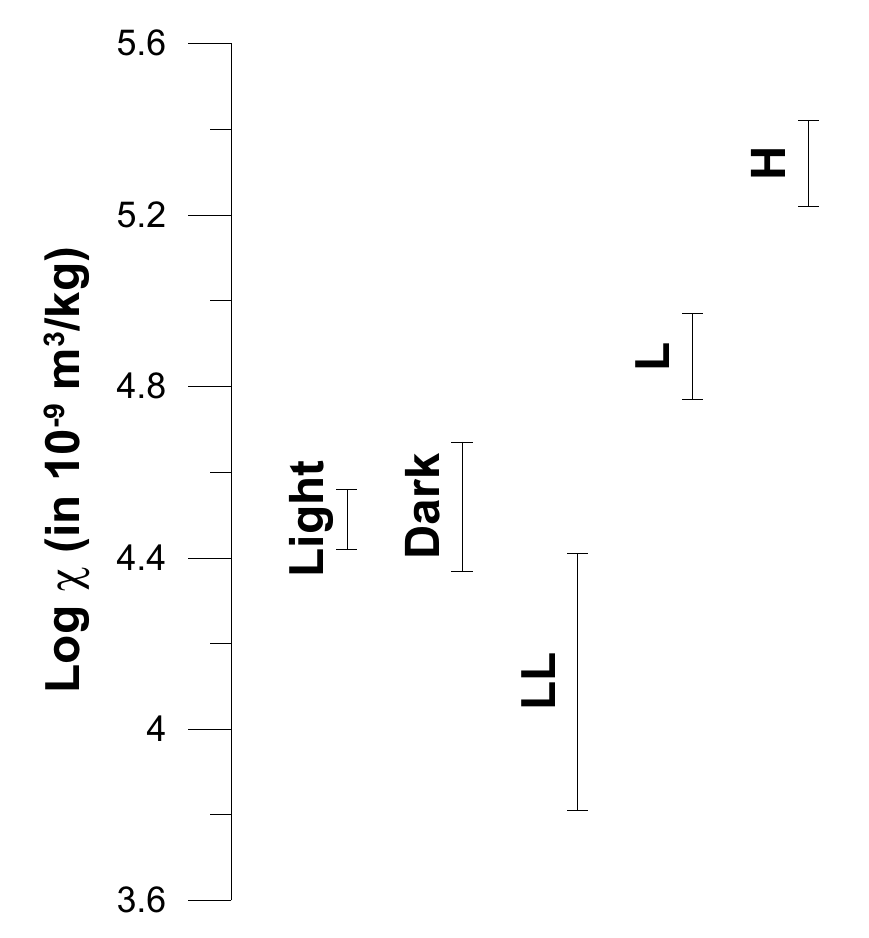}
\caption{Magnetic susceptibility of the Chelyabinsk light-colored and dark-colored lithologies compared to other H, L, and LL ordinary chondrites as reported in Rochette et al. \cite{rochette-2003}. The width of the error bars is equal to the standard deviation.}
\label{fig:seven}
\end{figure}

\begin{figure}[hbt]
\centering
\includegraphics{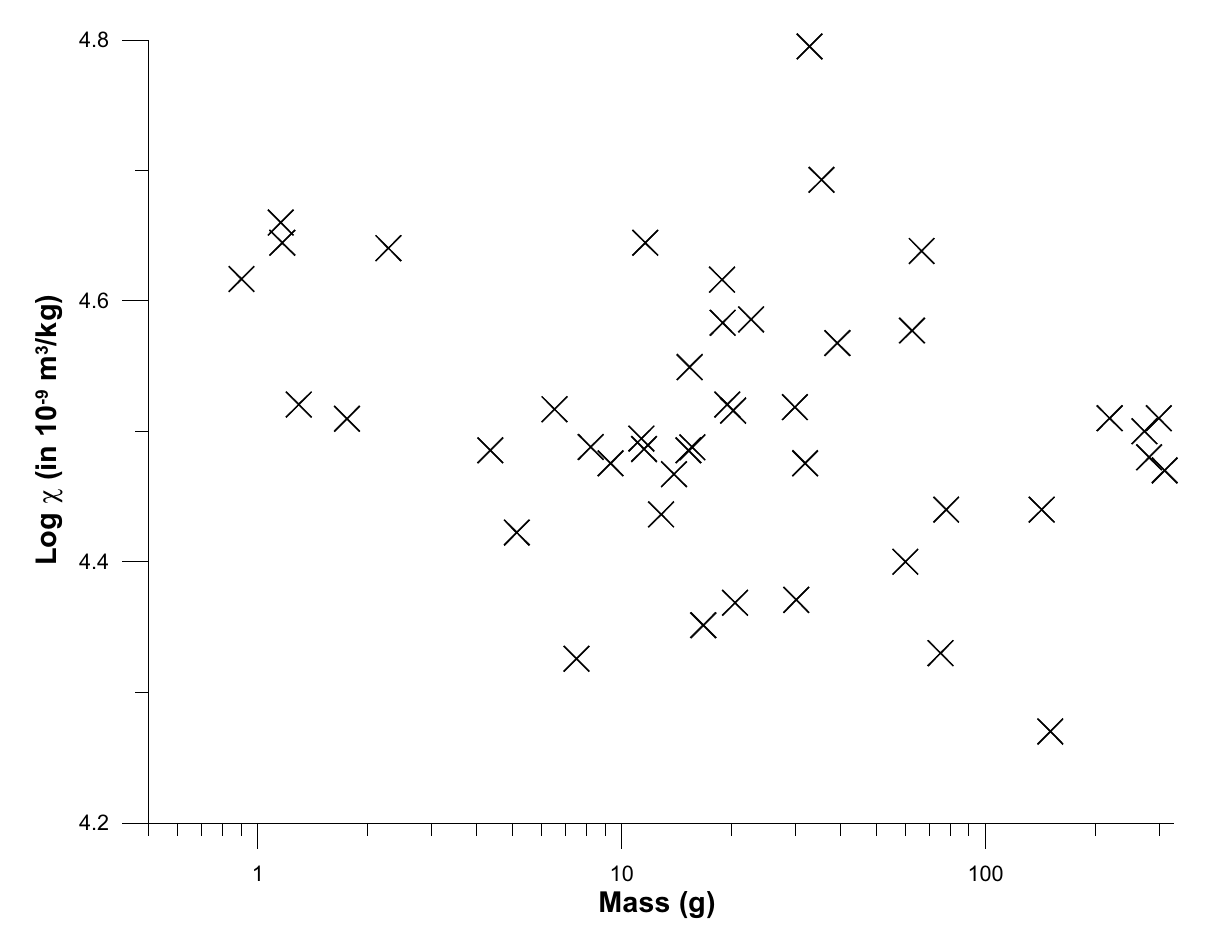}
\caption{Magnetic susceptibility (log $x$ in 10$^{-9}$ m$^3$/kg) of the Chelyabinsk meteorites as a function of their mass. No correlation is observed.}
\label{fig:eight}
\end{figure}

The results of the susceptibility measurements of the Chelyabinsk meteorites are summarized in Tables~\ref{tab:one} and \ref{tab:two}. The logarithm of the apparent magnetic susceptibility (in 10$^{-9}$ m$^3$/kg, Fig.~\ref{fig:seven}) ranges from 4.33 to 4.79 with the mean value of 4.49 s.d. 0.07 (light-colored lithology, 20 samples included) and 4.52 s.d. 0.15 (dark-colored, 8 samples included). This is within s.d. of the results from an independent study by Bezaeva et al. \cite{bezaeva-2013} on samples from the collections at the Vernadsky Institute, Russian Academy of Sciences, Moscow. However, the magnetic susceptibility values are higher compared to the other LL chondrite falls (average 4.10 s.d. 0.30) reported in Rochette et al. \cite{rochette-2003} and are closer to the transitional L/LL class (4.60 s.d. 0.11 \cite{rochette-2003}). The susceptibility values are quite uniformly distributed among samples of various masses (Fig.~\ref{fig:eight}). As magnetic susceptibility in ordinary chondrites mainly reflects the amount of ferromagnetic metallic iron, it is apparent that, compared to typical LL chondrites, the Chelyabinsk meteorites are enriched in metallic iron.

The dark-colored lithology shows slightly higher magnetic susceptibility (similarly to that reported by Bezaeva et al. \cite{bezaeva-2013}) and thus is slightly richer in metallic iron content. However, similarly to the density values, the light-colored and dark-colored material magnetic susceptibility ranges significantly overlap and susceptibility cannot be used as a reliable criterion to distinguish between the two lithologies.

\section*{Discussion}

The Chelyabinsk meteorite fragments represent fresh LL5 type chondrite material originating from a small, $\sim\!\!20$ m sized meteoroid (small asteroid). The meteorite samples are of petrographic type 5 and were subjected to recrystallization at approx. 700--750\degr C. This thermal metamorphism evidence suggests that their parent body must have been originally much larger. Therefore the meteoroid, which collided with the Earth and delivered the Chelyabinsk meteorites, was only a fragment of the LL chondrite parent body.

Moreover, the Chelyabinsk meteorites are composed of three lithologies: the light-colored, dark-colored, and impact-melt. The mineralogical analysis revealed that these lithologies are derived from the same LL5 chondritic source. The difference between them is in the shock. This gives us a unique opportunity to study effects of high shock on the mineralogy, reflectance spectra, and physical properties of chondritic material.

The light-colored lithology shows lower level of fracturing and the impact-melt lithology is localized to inter-granular space only. In contrast, the dark-colored lithology is shocked up to higher shock level causing higher level of fracturing. In addition to the presence of silicate impact-melt lithology similar to the one in the light-colored lithology, another metal and sulfide-rich melt population impregnating the inter- as well as intra-granular space is present. This metal and sulfide-rich melt is the darkening agent.

The silicate impact-melt lithology rather resembles the characteristics of whole-rock melt and it is present in both light- and dark-colored lithology meteorites. Some meteorites are even dominated by the impact-melt and thus it is distinguished as a third lithology type and should be not confused with the dark-colored lithology. In contrast, the metal and sulfide-rich melt present in dark-colored lithology is rather partial melt and appears uniformly distributed.

The higher shock level observed in the dark-colored lithology can be explained by a single shock event identical to the one in the light-colored lithology, only being stronger due to the darkened material being closer to the impact point. Alternatively, increased shock darkening can be caused by a secondary event affecting a part of the parent body only.

Both shock scenarios seem to be too strong to occur on the $\sim\!\!20$ m Chelyabinsk meteoroid. Thus, the shock most likely took place earlier on the parent body. It is possible that the last high-shock impact event responsible for the darkening and other shock features observed in both lithologies may have also caused the disintegration of the parent body and the origin of the Chelyabinsk meteoroid. Most likely, numerous smaller asteroid fragments similar to the Chelyabinsk meteoroid body were created during this event \cite{marcos-2013}.

The spectrum of the light-colored lithology is typical for an LL ordinary chondrite with the presence of 1 and 2 \micron{} olivine and pyroxene absorption bands. Such a spectrum is similar to fresh S or Q type asteroids. In contrast, the impact-melt and dark-colored (shock-darkened) lithology spectra are darker and the silicate absorption bands are almost invisible. This characteristic is similar to those reported earlier on other shock-darkened meteorites (e.g. \cite{britt-1989,britt-1989a,keil-1992,britt-1994}). Due to these characteristics it was suggested earlier (e.g. \cite{britt-1989b,keil-1992,britt-1994}) that shock-darkened chondritic material may be present on the surfaces of some dark asteroids (e.g. C type) and our results support this suggestion. Compared to impact melting, shock darkening seems to be a more efficient process in altering the reflectance spectra and causes small slope change. Shock also affects the albedo and silicate absorption bands of chondrite material in a way similar to space weathering. However, unlike space weathering (or at least unlike its lunar-type form with the presence of nanophase iron in the surface coatings of silicate minerals) there is negligible (impact-melt lithology) to minor (dark-colored lithology) change in the spectral slope observed.

The bulk and grain densities and the porosity of the light-colored and dark-colored lithologies are almost identical. The dark-colored lithology has a slightly smaller average grain density of 3.42 g/cm$^3$ compared to 3.51 g/cm$^3$ for the light-colored one. However, the slight difference between the two lithologies is smaller compared to the s.d. of these values. The dark-colored lithology is more fractured. Thus, one would expect a significantly lower bulk density and higher porosity. However, the fractures are filled by melt veins. This explains the observed negligible density differences between the light-colored and dark-colored lithologies. Similarly, the magnetic susceptibility logarithm is slightly higher (pointing on a slightly higher amount of metal related to the presence of troilite and metal-rich impact-melt veins) for the dark-colored lithology, but the difference is again within the s.d. and the susceptibility (as well as the density and porosity) values cannot be used to distinguish between the two lithologies. It was not possible to measure densities, porosity and magnetic susceptibility of impact-melt lithology because it appears only in association with light-colored and dark-colored lithology. However, values measured on meteorites with significant portion of impact-melt lithology (e.g. meteorite no. A2, Table~\ref{tab:one}) do not deviate from others. Thus, despite lithological variety in meteorite material, the Chelyabinsk meteorite shower is rather homogeneous in its physical properties, in resemblance to other meteorite showers reported in Consolmagno et al. \cite{consolmagno-2006}.

\section*{Conclusions}

The Chelyabinsk fireball with an almost instant recovery of a large number of fresh meteorites is one of the most spectacular meteorite fall events in the recent history. The recovered meteorite material was characterized as brecciated LL5 ordinary chondrite. Based on the magnetic susceptibility being in the intermediate range between LL and L chondrites, the Chelyabinsk meteorites are richer in metallic iron as compared to other LL chondrites. The measured bulk and grain densities and the porosity closely resemble other LL chondrites. Three lithologies, the light-colored, dark-colored, and impact-melt lithologies, were found within the recovered meteorites. The impact-melt is silicate-rich and is present within the light-colored and dark-colored lithology stones as dark inter-granular veins. In the dark-colored lithology, additional fine-grained metal and sulfide-rich melt forming a dense network of fine veins impregnating the inter- and intra-granular pore space is present. This metal and sulfide-rich melt is related to a higher shock experienced by the dark-colored lithology and is the main darkening agent.

The presence of the LL5 lithologies of distinct shock levels enables us to study shock-generated changes in the spectral and physical properties of chondritic material.

Both impact-melt generation and shock darkening cause a decrease in reflectance and a suppression of the silicate absorption bands in the reflectance spectra. Shock darkening seems to be a more efficient process in altering the reflectance spectra. Such spectral changes are similar to the space weathering effects observed on asteroids. However, space weathering of chondritic materials is often accompanied with significant spectral slope change (reddening). In our case, only negligible (impact-melt lithology) to minor (dark-colored lithology) change in the spectral slope is observed. Thus, it is possible that some dark asteroids with invisible silicate absorption bands may be composed of relatively fresh shock-darkened chondritic material. The main spectral difference of asteroid surfaces dominated by impact-melt or shock darkening, or space weathering, is a significant slope change (reddening) in the latter case.

Shock darkening does not have a significant effect on the material physical properties. The increased shock causes fracturing with expected porosity increase and bulk density decrease. However, this is compensated by impregnation of the fractures by fine-grained sulphide and metal-rich melt having an opposite effect on the porosity and bulk density. The presence of sulphide and metal-rich melt is responsible for the slight increase of the magnetic susceptibility observed in the dark-colored lithology.

\section*{Acknowledgements}

The work was supported by Academy of Finland Projects No. 257487 and No. 260027 (SIRONTA), Ministry of Education, Youth and Sports Czech Republic Grant LH12079, Federal Grant-In-Aid Program GC No. 14.740.11.1006, and the ERC Advanced Grant No. 320773 (SAEMPL). Authors would like to thank to Ilya Weinstein for help with the measurements arrangements, to Pasi Heikkil\"a, Alevtina Maksimova, Razilya Gizzatullina, Albina Zainullina, and Anastasia Uryvkova for help with the laboratory work, and to Hanna Pentik\"ainen for language corrections, and to to three anonymous reviewers for constructive comments on the manuscript.

\bibliographystyle{plainnat}

\begin{thebibliography}{99}

\bibitem{metbul}
Meteoritical Bulletin no. 102 (2014). \emph{Meteoritics \& Planetary Science} {\bf 48} (in preparation, also available on \url{http://www.lpi.usra.edu/meteor/metbull.php?code=57165}, read on 25.6.2013)

\bibitem{galimov-2013}
Galimov, E. M., Kolotov, V. P., Nazarov, M. A., Kostitsyn, Yu. A., Kubrakova, I. V., Kononkova, N. N., I. A. Roshchina, V. A. Alexeev, L. L. Kashkarov, D. D. Badyukov, and V. S. Sevast'yanov (2013). Analytical Results for the Material of the Chelyabinsk Meteorite. \emph{Geochemistry International} {\bf 51}:580--598. DOI: 10.1134/S0016702913070100

\bibitem{proud-2013}
Proud, S. R. (2013), Reconstructing the orbit of the Chelyabinsk meteor using satellite observations. \emph{Geophysical Research Letters} {\bf 40}:3351--3355. DOI: 10.1002/grl.50660

\bibitem{yeomans-2013}
Yeomans D. and Chodas P. (2013). Additional Details on the Large Fireball Event over Russia on Feb. 15, 2013. NASA/JPL Near-Earth Object Program, \url{http://neo.jpl.nasa.gov/news/fireball_130301.html}, read on 25.6.2013

\bibitem{kohout-2008}
Kohout, T., Kletetschka, G., Elbra, T., Adachi, T., Mikula, V., Pesonen, L.J., et al. (2008). Physical properties of meteorites --- applications in space missions to asteroids. \emph{Meteoritics \& Planetary Science} {\bf 43}:1009--1020. DOI: 10.1111/j.1945-5100.2008.tb00689.x

\bibitem{consolmagno-1998}
Consolmagno G. and Britt D. (1998). The density and porosity of meteorites from the Vatican collection. \emph{Meteoritics \& Planetary Science} {\bf 33}:1231--1241

\bibitem{macke-2010}
Macke R. J., Britt D. T., and Consolmagno G. J. (2010). Analysis of systematic error in ''bead method'' measurements of meteorite bulk volume and density. \emph{Planetary and Space Science} {\bf 58}:421--426. DOI: 10.1016/j.pss.2009.11.006

\bibitem{gattacceca-2004}
Gattacceca J., Eisenlohr P., and Rochette P. (2004). Calibration of in situ magnetic susceptibility measurements. \emph{Geophysical Journal International} {\bf 158}:42--49. DOI: 10.1111/j.1365-246X.2004.02297.x

\bibitem{rochette-2003}
Rochette P., Sagnotti L., Bourot-Denise M., Consolmagno G., Folco L., Gattacceca J., Osete M. L., and Pesonen L. J. (2003). Magnetic classification of stony meteorites: 1. Ordinary chondrites. \emph{Meteoritics \& Planetary Science} {\bf 38}:251--268

\bibitem{brearley-1998}
Brearley, A. J. and Jones R. H. (1998). Chondritic meteorites. In: Papike, J. J. (ed.). Planetary materials, Reviews in Mineralogy, 36, Mineralogical Society of America, 398 pages

\bibitem{stoffler-1991}
St\"offler D., Keil K., and Scott E. R. D. (1991). Shock metamorphism of ordinary chondrites. \emph{Geochimica Cosmochimica Acta} {\bf 55}:3845--3867

\bibitem{consolmagno-2008}
Consolmagno G., Britt D., and Macke R. (2008). The significance of meteorite density and porosity. \emph{Chemie der Erde} {\bf 68}:1--29. DOI: 10.1016/j.chemer.2008.01.003

\bibitem{bezaeva-2013}
Bezaeva, N. S., Badyukov, D. D., Nazarov, M. A., Rochette, P., and Feinberg, J. (2013). Magnetic Properties of the Chelyabinsk Meteorite: Preliminary Results. \emph{Geochemistry International} {\bf 51}:629--635. DOI: 10.1134/S0016702913070082

\bibitem{marcos-2013}
de la Fuente Marcos C. and de la Fuente Marcos R. (2013). The Chelyabinsk superbolide: a fragment of asteroid 2011 EO40? \emph{Monthly Notices of the Royal Astronomical Society} (in press). DOI: 10.1093/mnrasl/slt103

\bibitem{britt-1989}
Britt, D. T. and Pieters, C. M. (1989). Bidirectional Reflectance Characteristics of Black Chondrite Meteorites. Abstracts of the Lunar and Planetary Science Conference 20:109--110

\bibitem{britt-1989a}
Britt, D. T., Pieters, C. M., Petaev, M. I., and Zaslavaskaya, N. I. (1989a). The Tsarev meteorite --- Petrology and bidirectional reflectance spectra of a shock-blackened L chondrite. In: Lunar and Planetary Science Conference, 19th, Houston, TX, Mar. 14--18, 1988, Proceedings (A89-36486 15--91). Cambridge/Houston, TX, Cambridge University Press/Lunar and Planetary Institute, 1989, p. 537--545

\bibitem{keil-1992}
Keil, K., Jeffrey, F. B., and Britt D. T. (1992). Reflection spectra of shocked ordinary chondrites and their relationship to asteroids. \emph{Icarus} {\bf 98}:43--53. DOI: 10.1016/0019-1035(92)90205-L

\bibitem{britt-1994}
Britt, D. T. and Pieters, C. M. (1994). Darkening in black and gas-rich ordinary chondrites: The spectral effects of opaque morphology and distribution. \emph{Geochimica et Cosmochimica Acta} {\bf 58}:3905--3919. DOI: 10.1016/0016-7037(94)90370-0

\bibitem{britt-1989b}
Britt, D. T., Pieters, C. M., Webb, R. S., and Pratt, S. F. (1989b). Relationship of C-type Asteroids to Dark Meteorites: Evidence for Optical Alteration by Asteroidal Regolith Processes. Abstracts of the Lunar and Planetary Science Conference 20:111--112

\bibitem{consolmagno-2006}
Consolmagno G. J., Macke R. J., Rochette P., Britt D. T., and Gattacceca J. (2006). Density, magnetic susceptibility, and the characterization of ordinary chondrite falls and showers. \emph{Meteoritics \& Planetary Science} {\bf 41}:331--342


\end{thebibliography}

\end{document}